\documentclass[12pt]{iopart}
\usepackage{iopams}
\eqnobysec
\newcommand{\be}{\begin{equation}}
\newcommand{\ee}{\end{equation}}

\newcommand{\ba}{\begin{array}}
\newcommand{\ea}{\end{array}}
\newcommand{\bea}{\begin{eqnarray}}
\newcommand{\eea}{\end{eqnarray}}

\newcommand{\nn}{\nonumber \\}
\newcommand{\bel}{\begin{equation}\label}

\begin{document}

\title[Exceptional Superintegrability]{Families of superintegrable Hamiltonians constructed from exceptional polynomials}
\author{Sarah Post$^1$,  Satoshi Tsujimoto$^2,$ and Luc Vinet$^{1}$  }
\address{$^1$ Centre de Recherches Math\'ematiques. Universit\'e de Montr\'eal. Montr\'eal CP6128 (QC) H3C 3J7, Canada}
\address{$^2$ Department of Applied Mathematics and Physics.
Graduate School of Informatics.
Kyoto University.
Yoshida-Honmachi, Sakyo-ku, Kyoto
606-8501, Japan}
\ead{post@crm.umontreal.ca, luc.vinet@umontreal.ca}
\begin{abstract} 
We introduce a family of exactly-solvable two-dimensional Hamiltonians whose wave functions are given in terms of Laguerre and exceptional Jacobi polynomials. The Hamiltonians contain purely quantum terms which vanish in the classical limit leaving only a previously known family of superintegrable systems. 
Additional, higher-order integrals of motion are constructed from ladder operators for the considered orthogonal polynomials proving the quantum system to be superintegrable.

\end{abstract}
\pacs{ 02.30.Gp, 02.30.Hq,  03.65.Fd, 03.65.Ge,  12.60.Jv }
\ams{15A18, 05E35, 33D45, 34Kxx, 81Q60 }

\section{Introduction}
The connection between classical families of orthogonal polynomials and exactly-solvable, integrable and superintegrable systems is well-known. In this paper, we show that the same connection applies to the recently discovered families of exceptional orthogonal polynomials \cite{gomez2009extended}. 
In particular, we demonstrate the existence of an infinite family of Hamiltonians which are both superintegrable and exactly-solvable and whose wavefunctions are composed of the product of classical Laguerre and exceptional Jacobi polynomials. Furthermore, in the classical limit, the Hamiltonian reduces to the celebrated Tremblay-Turbiner-Winternitz system \cite{TTW2009, TTW2010}. 

A system is said to be superintegrable if it admits more integrals of motion than degrees of freedom. For the purpose of this article, we consider a Hamiltonian on the two-dimensional Euclidean plane 
\be H=\frac{1}{2}(p_1^2+p_2^2)+V(x_1,x_2)\ee
which admits two addition integrals of motion
\be \label{2}L_a=\sum_{0\leq j+k\leq n}f_{a, j k}(x_1,x_2)p_1^jp_2^k, \qquad a=1,2,\ee
where the $p_i$ are the components of the momenta conjugate to $x_i$ and are taken to be, in the quantum system, 
\be p_i=-i\hbar\frac{\partial}{\partial x_i}.\ee
As indicated in \eref{2}, the additional integrals will be assumed to be polynomial in the momenta and the degree of the system is said to be $n$, the highest order of the integrals. In classical mechanics the three functions $\{ L_1, L_2, H\}$ are assumed to be functionally independent and in quantum mechanics they are assumed to be algebraically independent. While the integrals, $L_a,$ Lie or Poisson commute with the Hamiltonian $H$, they do not commute with each other and so generate an algebra which usually closes to form a polynomial algebra \cite{granovskii1991quadratic, letourneau1995superintegrable, Dask2001, marquette2009painleve}. 

The study of superintegrable systems began with second-order superintegrability \cite{FMSUW, WSUF, MSVW1967, Evans1990, SCQS}, including the best-known examples of the harmonic oscillator \cite{JauchHill1940, MoshSmir} and the Kepler-Coulomb system \cite{Fock,Bargmann}. Recently, new families of superintegrable systems have been discovered with integrals of arbitrary order. The first was discovered by Tremblay, Turbiner and Winternitz \cite{TTW2009, TTW2010} 
\be\label{HTTW} H^{TTW}=\frac{1}{2}\left(p_r^2+\frac{1}{r^2}p_\theta^2\right)+\omega^2 r^2+\frac{k^2}{r^2}\left(\frac{a}{\cos^2(k\theta)}+\frac{b}{\sin^{2}(k\theta)}\right).\ee
The discovery of these new families has lead to much new research on the discovery and treatment of such superintegrable systems, see e.g. \cite{KKM2010JPA, PW20101, QuesneTTWodd, KKM2010quant}. 

The connection between superintegrable systems and orthogonal polynomials is most obviously evident in the conjecture that all superintegrable systems are exactly-solvable \cite{TempTW}. Recall, a system is said to be exactly solvable if its energy values can be calculated algebraically and the wave functions can be written in terms of orthogonal polynomials multiplied by the ground state 
\cite{Turbiner1988, gonzalez1994quasi, lapointe1996exact}. This connection was recently exploited  by Kalnins, Kress and  Miller who made use of ladder operators for orthogonal polynomials to construct additional integrals of motion, thus proving the superintegrability of several families of superintegrable systems, including the TTW system \cite{KKM2011Recurr}. This method has been successfully been applied to families of superintegrable systems with reflections in the potential \cite{PVZ2011inffam} as well as scalar potentials defined on pseudo-Euclidean space \cite{LPW2012}. Ladder operators associated with shape invariance have also been used to construct infinite families of superintegrable systems \cite{marquette2009super, marquette2010superintegrability, marquette2011infinite}.

The purpose of this article is to extend this analysis to exactly-solvable systems whose wave functions are expressible in terms of exception orthogonal polynomials. Exceptional orthogonal polynomials are eigenfunctions of Sturm-Liouville equations which generalize the classical families of orthogonal polynomials in the sense that, unlike the case of classical polynomials, the families of exceptional orthogonal polynomials admit gaps in their degree \cite{gomez2009extended,GKMreview}. The Sturm-Liouville equations associated with such polynomials have been recently the subject of intense research including their connection with shape invariance, see e.g. \cite{quesne2008exceptional, odake2009infinitely,ho2009properties, odake2010another, quesne2011higher, odake2011new}, and their connection with  Darboux-Crumb transformations, see e.g. \cite{sasaki2010exceptional, gomez2011two, gomez2012conjecture}. 

The plan of the paper is as follows. In section 2, a new exactly-solvable two-dimensional Hamiltonian is introduced and its wavefunctions and spectrum are found as well as its classical limit. In section 3 the Hamiltonian is proven to be superintegrable for rational $k$ by direct construction of the integrals of motion. Section 4 gives the $k=1$ example and explores the connection with its classical limit. Section 5 is comprised of concluding remarks.

\section{An exactly-solvable Hamiltonian}
Consider a generalization of the TTW Hamiltonian \eref{HTTW} given in polar coordinates by  
\be\label{Hk}\fl H_k=-\frac{1}2\Delta +\frac12 \omega^2 r^2 +\frac{k^2}{2r^2}\left(\frac{\alpha^2-\frac14}{\sin^2(k\phi)}+\frac{\beta^2-\frac14}{\cos^2(k\phi)} +\frac{4\left(1+b\cos(2k\phi)\right)}{\left(b+\cos(2k\phi)\right)^2}\right)\ee
where 
\[ b=\frac{\beta+\alpha}{\beta-\alpha}.\]
The Schr\"odinger equation associated with this Hamiltonian \eref{Hk} 
\be \label{Schro} H_k\Psi-E\Psi=0.\ee
separates in polar coordinates as 
\be \Psi =\Phi(\phi) R(r)\ee
with 
\be\label{rad} \left(-\frac1{2r}\partial_r r\partial_r+\frac12\omega r^2+\frac{k^2A^2}{2r^2}-E\right)R(r)=0\ee
 \be \label{ang} \left(-\frac{1}{k^2}\partial_\phi^2+\frac{\alpha^2-\frac14}{\sin^2(k\phi)}+\frac{\beta^2-\frac14}{\cos^2(k\phi)} +\frac{4\left(1+b\cos(2k\phi)\right)}{\left(b+\cos(2k\phi)\right)^2}-A^2\right)\Phi(\phi)=0.\ee
Note that the radial equation is exactly that of a two-dimensional oscillator and the angular part is a deformation of a Darboux-Poschl-Teller potential \cite{darboux1887leccons, poschl1933bemerkungen}. 
 
 In fact, the angular part of the Hamiltonian \eref{Hk} is the shape-invariant Hamiltonian introduced in \cite{sasaki2010exceptional} whose eigenfunctions can be written in terms of the $X_1$ Jacobi polynomials. The change of variables 
 \be \Phi(\phi)=X_n(x), \qquad x=\cos(2k\phi),\qquad  n\geq 1,\ee
transforms \eref{ang} into 
 \be \label{angg}  \left(G_xT^{\alpha, \beta} G_x^{-1} -A^2\right)X_n(x)=0\ee
 where 
 \bea\label{Gx} G_x=\frac{(1-x)^{\frac{\alpha}2+\frac14}(1+x)^{\frac{\beta}2+\frac14}}{(x-b)}\\
 \fl \label{Tab} T^{\alpha, \beta} =4(x^2-1)\partial_x^2+\frac{4(\beta-\alpha)(1-bx)}{b-x}\left((x+b)\partial_x -1\right)+(\alpha+\beta+1)^2.\eea
 The operator $T^{\alpha, \beta}$ is equivalent to the eigenvalue operator for the $X_1$ exceptional Jacobi polynomials \cite{gomez2009extended}.  The eigenvectors of \eref{Tab} are given by the exceptional Jacobi polynomials $\widehat{P_n}^{\alpha, \beta}$,
 \be \fl\label{Phat} \widehat{P_n}^{\alpha, \beta}=-\frac12(x-b)P_{n-1}^{\alpha, \beta}(x)+\frac{1}{2n-2+\alpha+\beta} \left[bP_{n-1}^{\alpha, \beta}(x)-P_{n-2}^{\alpha, \beta} (x)\right], \qquad n\geq 1,\ee
 where $P_{n}^{\alpha, \beta}(x)$ are the standard Jacobi polynomials, \cite{sasaki2010exceptional}. 
 The eigenvalue equation \eref{angg} has eigenvalue 
 \be\label{An} A^2\equiv A_n^2=(2n-1+\alpha+\beta)^2, \qquad n\geq 1.\ee
  
  For \eref{rad}, the change of variables
 \be \label{Y} R(r)=Y_m^{A_n}(y), \qquad y=\omega r^2\ee and conjugation by the ground state 
 \be G_y=y^{A_n/2}e^{-y/2},\ee
transforms \eref{rad} to 
 \be \label{rady} y\partial_y^2Y+(1+k A_n-y)\partial_y Y+\frac{E}{4\omega}Y=0.\ee
 The solutions of \eref{rady} are given in terms of Laguerre polynomials
 \be Y_m^{A_n}=G_y L_m^{A_n}(y),\ee
 whenever the energy is quantized as
 \be\label{Emn} E_{m,n}=\omega\left(2m +k(2n+\alpha+\beta-1)+1\right).\ee
Unless needed to avoid confusion, the subscripts on $A$ and $E$ will be dropped in what follows. 
 
 \subsection{The classical limit}
 It is interesting to observe that in the classical limit, this system reduces to the TTW system \cite{TTW2009}. In fact, we re-introduce the parameter $\hbar$ by multiplying the system by a factor of $\hbar^2$. The corresponding Schr\"odinger equation is equivalent to the $\hbar=1$ case, with an appropriate scaling of the energy. The Hamiltonian becomes
 \be \label{Hkhbar}\fl H_k=-\hbar^2\frac{1}2\Delta +\frac12 \hbar^2\omega^2 r^2 +\frac{\hbar^2 k^2}{2r^2}\left(\frac{\alpha^2-\frac14}{\sin^2(k\phi)}+\frac{\beta^2-\frac14}{\cos^2(k\phi)} +\frac{4\left(1+b\cos(2k\phi)\right)}{\left(b+\cos(2k\phi)\right)^2}\right).\ee
 We renormalized the parameters so that the potential is not annihilated in the classical limit ($\hbar\rightarrow 0$) by taking 
 \be \widehat{\alpha} =\hbar \alpha, \qquad \widehat{\beta} =\hbar \beta,\qquad \widehat{\omega} =\hbar \omega. \ee
 The new Hamiltonian becomes
 
 \be \label{Hkhbarhat}\fl H_k=-\hbar^2\frac{1}2\Delta +\frac12 \widehat{\omega}^2 r^2 +\frac{ k^2}{2r^2}\left(\frac{\widehat{\alpha}^2-\frac{\hbar^2}4}{\sin^2(k\phi)}+\frac{\widehat{\beta}^2-\frac{\hbar^2}4}{\cos^2(k\phi)} +\frac{4\hbar^2\left(1+\widehat{b}\cos(2k\phi)\right)}{\left(\widehat{b}+\cos(2k\phi)\right)^2}\right),\ee
 where 
 \be \widehat{b}=\frac{\widehat{\beta}+\widehat{\alpha}}{\widehat{\beta}-\widehat{\alpha}}=b, \qquad \lim_{\hbar\rightarrow 0}\widehat{b}=\widehat{b}.\ee
 In this form, it is easy to see that the classical limit of the system \eref{Hkhbarhat} is exactly the TTW system and respects the requirement for bounded trajectories. Namely, in this notation, all bounded trajectories satisfy $\widehat{\omega}^2>0, \, \widehat{\alpha}^2>0, \, \widehat{\beta}^2>0$.

 \section{Superintegrability}
 It is immediately obvious that the Hamiltonian \eref{Hk} is integrable with first integral associated with separation of variables and given by 
 \be L_1=-\frac{1}{k^2}\frac{\partial^2}{\partial \theta^2} +\left(\frac{\alpha^2-\frac14}{\sin^2(k\phi)}+\frac{\beta^2-\frac14}{\cos^2(k\phi)} +\frac{4\left(1+b\cos(2k\phi)\right)}{\left(b+\cos(2k\phi)\right)^2}\right).\ee
In this section, we will show that the Hamiltonian is also superintegrable by constructing two additional integrals of motion for the Hamiltonian \eref{Hk} using ladder operators for the orthogonal polynomials of the wavefunctions, as in \cite{KKM2011Recurr}.

The key to the method is to  utilize ladder operators which transform the wave functions but leave the energy fixed, i.e. automorphisms on the energy eigenspaces.   To this end, assume $k=p/q$, then the transformations 
\bea \label{mp}  m \rightarrow m+p, \qquad n\rightarrow n-q\\ m\rightarrow m-p, \qquad n\rightarrow n+q, \eea
do not change the energy \eref{Emn}.

 \subsection{Ladder operators for the exceptional Jacobi polynomials}\label{ladderexc}
 Ladder operators for the exceptional Jacobi polynomials can be constructed from ladder operators for the Jacobi polynomials (see e.g.\cite{koekoek1996askey})
 \bea\label{Ln}\fl  \mathcal{L}_n=\frac{(1-x^2)(2n+\alpha+\beta)}{2}\partial_x-\frac{n\left((2n+\alpha+\beta)x+\alpha-\beta+2\right)}{2}\\
\label{Rn}\fl \mathcal{R}_n=-\frac{(1-x)(2n+\alpha+\beta+2)}{2}\partial_x+\frac{(n+\alpha+\beta+1)\left((2n+\alpha+\beta)x+\alpha-\beta+2\right)}{2}
\eea
 and the "forward" and "backward" operators for the exceptional Jacobi polynomials \cite{ho2009properties}
 \bea \label{F} \mathcal{F}=\left(x-1\right)\left(x+\frac{\alpha+\beta}{\alpha-\beta}\right)\partial_x +\left(\alpha-1\right) t \left(x+\frac{2+\alpha+\beta}{\alpha-\beta}\right)\\
\mathcal{B}=-\frac{\alpha-\beta}{\alpha+\beta-(\alpha-\beta)x}\left((1+x)\partial_x+\beta\right), \label{B}\eea
whose actions are
\bea\mathcal{F} P_n^{\alpha+1, \beta-1}(x)=(2n-2+2\alpha)\widehat{P}_{n+1}^{\alpha, \beta}(x)\nn
\mathcal{B} \widehat{P}_{n+1}^{\alpha, \beta}(x)=\frac{-1}2 (n+\beta+1) P_{n}^{\alpha+1, \beta-1} (x)\nonumber.\eea

 The corresponding ladder operators for the exceptional Jacobi polynomials are defined as
 \be \mathcal{L}_{1,n}\equiv F\circ \mathcal{L}_n \circ B , \qquad \mathcal{R}_{1,n}\equiv F\circ \mathcal{R}_n \circ B.\ee
Conjugating by the ground state, $G_x$ as in \eref{Gx}, gives ladder operator for the angular component of the wave function
\bea J_{-,n}\equiv G_x\mathcal{L}_{1,n} G_x^{-1}, \qquad  J_{+,n}\equiv G_x\mathcal{R}_{1,n} G_x^{-1},\\
J_{-,n} X_{n}(x)=-(n+\alpha)(n+\alpha-2)(n+\beta)(n+\beta-2)X_{n-1}(x)\\
J_{+,n}X_{n}(x)=-(n)(n+\beta)(n+\alpha)(n+\alpha+\beta)X_{n+1}(x).\eea
The repeated application of the operators $J_{\pm, n}$ are defined as 
\be J_{\pm, n}^q=J_{\pm, n \pm (q-1)}\circ \ldots \circ J_{\pm,n \pm 1} \circ J_{\pm, n},\ee
with action on the basis as 
\bea \fl\qquad \ba {l}J_{-,n}^qX_{n}=(-1)^q(-n-\alpha)_q(-n-\alpha+2)_q(-n-\beta)_q(-n-\beta+2)_qX_{n-q}\\
        J_{+,n}^qX_{n}=(-1)^q(n)_q(n+\beta)_q(n+\alpha)_q(n+\alpha+\beta)_qX_{n+q}.\ea \label{recurJ}\eea

 \subsection{Ladder operators for the Laguerre polynomials}\label{ladderlag}
 Based on the ladder operators for the Laguerre polynomials, again see e.g. \cite{koekoek1996askey},   ladder operators for the functions $Y_m^{kA}(y)$ can be constructed as
\bea  K_{\pm kA,E}Y_m^{kA}(y)&=\left[ (1\pm kA)\partial_y-\frac{E}{4\omega}\mp\frac{kA}{2y}(1+kA)\right]Y_m^{k A}(y)\\
 &= k_{\pm} Y_{m\mp 1}^{kA\pm 2}(y),\nonumber\eea
where
\bea k_{+}=-1, \qquad k_{-}=-(m+1)(m+kA),\eea
and $E$ and $A$ take the quantized values as above \eref{Emn}, \eref{An}
\[ E=2\omega[2m+kA +1], \qquad A=2n+\alpha+\beta-1\]
To obtain the desired shift in the quantum numbers \eref{mp}, the p-fold composition of $K$ is defined  with the corresponding value of $kA$ shifted in each successive application
\bea K^{p}_{kA,E}\equiv K_{kA+2(p-1),E}\cdots K_{kA+2,E} K_{kA, E}\\
     K^{p}_{-kA,E}\equiv K_{-(kA-2(p-1)),E}\cdots K_{-(kA-2),E} K_{-kA,E}.\eea
The action of these operators on the wave functions is given by 
\bea \ba{ll} K^{p}_{kA,E}Y_{m}^{kA}&=(-1)^pY_{m-p}^{kA+2p}\\
     K^{p}_{-kA,E}Y_{m}^{kA}&=(-1)^p(m+1)_p(kA+m-p+1)_p Y_{m+p}^{kA-2p}.\ea \label{recurK}\eea
It is important to note that although the quantity $E$ is a function of $m$ and $n$, it is unchanged by the operation $m\rightarrow m\pm 1, $ and $kA \rightarrow kA \mp 2.$ Hence, the energy $E$ remains fixed in each successive applications of the operator.

 \subsection{Quantum-number independent integrals of motion}
Combing the two sets of ladder operators in subsections \ref{ladderexc} and \ref{ladderlag}, operators can be constructed which transform within fixed energy eigenspaces. The corresponding operators are 
 \be \Xi_{+}=  K^{p}_{kA,E}J_{+,n}^q, \qquad  \Xi_{-}=  K^{p}_{-kA,E}J_{-,n}^q,\ee
 which depend on the quantum numbers $m $ and $n$ and fix the energy $E$, so in fact
\[ [\Xi_{\pm}, H]\Psi_{m,n}=0.\] 
 In order to obtain differential operators which commute with the Hamiltonian for all values of $m$ and $n$, these quantum numbers must be removed from the operators. To do this, first, the energy $E$ is removed by moving the constant $E$ to the right and the replacing it with $H$. See the derivation in the appendix of  \cite{PVZ2011inffam}  for an explicit representation of these operators.  To remove the quantum number $n$,  note that under the transformation $n\rightarrow -n-\alpha-\beta+1$, the operators $\mathcal{L}_n$ and $\mathcal{R}_n$ \eref{Ln},\eref{Rn} transform as 
 \be \mathcal{L}_{n}\rightarrow \mathcal{R}_n, \qquad \mathcal{L}_{n+\ell}\rightarrow \mathcal{R}_{n-\ell}.\ee
This action transfers to the operators $\mathcal{L}_{1,n}$ and $\mathcal{R}_{1,n}.$ Similarly, the action  $n\rightarrow -n-\alpha-\beta+1$ sends $A\rightarrow -A$ and so interchanges the raising and lowering operators for the Laguerre polynomials, after the energy $E$ has been replaced by $H$. Thus, the operators
\be \Xi_2=\Xi_{+}+\Xi_{-}, \qquad \Xi_3=\frac{1}{2n+\alpha+\beta-1}\left(\Xi_{+}-\Xi_{-}\right),\ee
 are polynomial in $A^2$ and hence the following operators 
 \be  L_2=\left(\left(\Xi_{2}\right)_{E=H}\right)_{A^2=\frac{L_1}{k^2}}, \qquad L_3=\left(\left(\Xi_{3}\right)_{E=H}\right)_{A^2=\frac{L_1}{k^2}}\ee
 are independent of the quantum numbers $m$ and $n$ and commute with the Hamiltonian $H$. Thus, $L_2$ and $L_3$ commute with the Hamiltonian on the basis and hence commute as operators based on a combination of writing the commutator,  $[L_i, H]$, as a second-order operator with coefficients that depend on $H$ and $L_1$ \cite{KKM2010quant} and using a standard Wronskian argument about the separated solutions \cite{KKM2011Recurr}. 
 
 The algebra relations of the integrals can be obtained directly from the expansion coefficients \eref{recurJ}, \eref{recurK}, including the fact that neither $L_2$ nor $L_3$ commutes with  $L_1$ so the operators are algebraically independent.

\section{Conclusions}
 In this paper, we have introduced an exactly-solvable system whose wavefunctions are given in terms of a product of Laguerre and exceptional Jacobi polynomials. By construction, the Hamiltonian \eref{Hk} is integrable and, in addition, it admits higher, independent integrals of motion making it superintegrable.  These higher-order integrals are constructed directly from ladder operators for the considered orthogonal polynomials. In particular, the ladder operators for the exceptional Jacobi polynomials are constructed from the "forward" and "backward" operators \eref{F}, \eref{B} composed with the ladder operators for the classical Jacobi polynomials \eref{Ln}, \eref{Rn}. 
 
 It should be noted than this method can be directly applied to prove the superintegrability of the following Hamiltonian, with a Kepler-Coulomb type potential in the radial variable, 
 \be\fl H=-\frac{1}2\Delta +\frac{K}{r}+\frac{k^2}{2r^2}\left(\frac{\alpha^2-\frac14}{\sin^2(k\phi)}+\frac{\beta^2-\frac14}{\cos^2(k\phi)} +\frac{4\left(1+b\cos(2k\phi)\right)}{\left(b+\cos(2k\phi)\right)^2}\right).\ee
 In fact, this Hamiltonian is related to the Hamiltonian \eref{Hk} via coupling constant metamorphosis, see \cite{PW20101, HierGrammStackel, KMPostStackel}.
 
This method of constructing Hamiltonians and their integrals of motion can be extended in a straightforward manner to other families of exceptional polynomials. Most immediately, the angular part of the Hamiltonian given above  \eref{Hk} can be replaced by any of the infinite families of one-dimensional Hamiltonians for the $X_\ell$ exceptional polynomials \cite{odake2009infinitely, odake2010another}. Additionally, other families of Hamiltonians, say separable in Cartesian coordinates, can be obtained in a similar way from the Sturm-Liouville equations for other exceptional polynomials, e.g. extensions of the singular harmonic oscillator via exceptional Laguerre polynomials. These systems will be treated in future work.

\section*{References}
\bibliography{all}
 \bibliographystyle{unsrt}

\end{document}